\documentclass[aps,preprint,amssymb,12pt,floatfix]{revtex4}
\usepackage{subfigure}
\usepackage{graphicx}
\usepackage{rotating}

\begin{document}

\title{Probing the Mechanisms of Fibril Formation Using Lattice Models.}

\author{Mai Suan Li$^{1}$, D. K. Klimov$^2$, J. E. Straub$^3$, 
and D. Thirumalai$^{4,5}$}

\address{$^1$Institute of Physics, Polish Academy of Sciences,
Al. Lotnikow 32/46, 02-668 Warsaw, Poland\\
$^2$Bioinformatics and Computational Biology Program, School of Computational Sciences,
George Manson University, Manassas, VA 20110\\
$^3$ Department of Chemistry,
Boston University, MA 02215\\
$^4$Department of Chemistry and Biochemistry, University of Maryland,
College Park, MD 20742\\
$^5$Biophysics Program, Institute for Physical
Science and Technology, University of Maryland, College Park, MD 20742
}

\date{\small \today}

\baselineskip = 22pt

\begin{abstract}
Using exhaustive Monte Carlo simulations we study the kinetics and
mechanism of fibril formation using lattice models as a function of
temperature and
the number of chains.  While these models are, at best,
caricatures of peptides,
we show that a number of generic features thought to govern fibril
assembly are present in the toy model. The 
monomer, which contains eight beads made from three
letters (hydrophobic, polar, and charged), adopts
a compact conformation in the native state.
The kinetics of fibril assembly occurs in three distinct stages. In
each stage
there
is a cascade of events that transforms the monomers and
oligomers to ordered structures. In the
first "burst" stage highly mobile oligomers of varying sizes form. The
conversion to the aggregation-prone conformation occurs within the
oligomers during the second stage.
As time progresses, a dominant cluster emerges that contains a majority of
the chains. In the final stage,
the aggregation-prone conformation particles serve as a template onto
which smaller oligomers or monomers can dock and undergo conversion to
fibril structures. The overall time for growth in the latter stages is
well described by the Lifshitz-Slyazov growth kinetics for
crystallization
from super-saturated solutions. 
\end{abstract}


\maketitle

\section{Introduction}

The link between aggregation of proteins and a number
 of neurodegenerative diseases
has spurred many experimental
\cite{Lomakin96PNAS,Rochet00COSB,Wetzel02Structure,Selkoe03Nature,Dobson04Science,Ross04NatMed,Bossy-Wetzel04NatMed,Lee05JACS,Nelson06APC,Chiti_ARBioChem06}
and 
theoretical studies \cite{Gupta98ProtSci,Ma02PNAS,Massi01Proteins,Smith01JMB,Gsponer03PNAS,Klimov03Structure,Thirumalai03COSB,Farvin04BiophysJ,Wei04BJ,Buchete05JMB,Takeda07JMB,Bellesia07JCP,Baumketner07JMB}.
Aggregation rates depend not only
on protein sequence but also on the concentration
of proteins and external conditions (temperature, pH, presence of crowding
agents etc.). The observation that many proteins that are unrelated
by sequence and
structure can aggregate and form fibrils \cite{Chiti_ARBioChem06} with
similar morphologies (albeit under different growth condition)
suggests that certain generic aspects of
oligomerization and subsequent fibril growth can be 
gleaned from toy models.
Towards this end a number of lattice models
\cite{Gupta98ProtSci,Harrison01ProtSci,Dima02ProtSci}
have been introduced to probe the fibril formation mechanism.
Here, following the important studies by Hall and coworkers
\cite{Gupta98ProtSci,Teplow06ACR}, we use a 
three-dimensional lattice model that is, in part, inspired by all-atom
simulations of
oligomer formation of the peptide fragment $A\beta _{16-22}$
\cite{Klimov03Structure}, to
provide insights into mechanism of  fibril formation.


Soluble ({\bf S}) monomeric polypeptide chain can be either random coil-like
(A$\beta$ peptides or $\alpha$-synuclein) or
folded (transthyretin).
 Typically, fluctuations or denaturation stress can populate
one of several aggregation-prone
conformations ({\bf N$^*$}).
Because of conformational variations in
{\bf N$^*$} fibrils with differing molecular structure
can form starting from the same sequence.
However, the growth mechanism starting from {\bf N$^*$} to the fibril state
is not fully understood.
Three mechanisms
for fibril assembly have been proposed. In the nucleation-growth (NG) mechanism
\cite{Jarrett93Cell} the first step is the oligomerization of sufficient number of {\bf N$^*$} particles
oligomerize and form a critical nucleus, which is a free-energetically an 
uphill process upon forming {\bf N$^*$}$_n$ ($n >$ the size of the
critical nucleus $n_c$). {\bf S} monomers can rapidly add to the oligomer 
resulting in growth of oligomers and eventual fibril assembly. The
templated-assembly (TA) process 
\cite{Esler00Biochem,Cannon04AnalBiochem,Nguyen07PNAS}
suggests that preformed {\bf N$^*$}$_n$ complex, with presumably
 $n > n_c$, serves as a
template onto which {\bf S} or {\bf N$^*$} can dock and undergo the needed
structural arrangement to lock onto the template. Based on kinetic data 
on prion formation in yeasts the nucleated conformational conversion (NCC) model
\cite{Lomakin96PNAS,Serio00Science} has been proposed. In the NCC
model it is envisioned that {\bf S} forms mobile disordered oligomers. The monomers
in the oligomer undergo {\bf S} $\rightarrow$ {\bf N$^*$} conversion
 to form nuclei
{\bf N$^*$}$_n$. The species {\bf N$^*$}$_n$ can serve as a template and incorporate
other (less structured) oligomers or monomers to rapidly form ordered
fibrils. The important feature of NCC is that structural arrangement
{\bf S} $ \rightarrow$ {\bf N$^*$} $\rightarrow N_{FIB}$ ($N_{FIB}$ is the monomer structure
in the fibril) occurs within the molten oligomer.
In many cases the structures of {\bf N$^*$} and $N_{FIB}$ are similar.

In this paper we study 
the mechanism of fibril assembly using a simple lattice model for
which extensive simulations can be performed.
The analysis reveals a complex scenario for protofilament and fibril
assembly that seems to
 have elements of all the three growth models.
The dependence of fibril
formation time $\tau _{fib}$ on the number of monomers
reveals that late stages of growth have a lot in common with crystallization
in super saturated solutions.
These findings arise from detailed Monte Carlo (MC)
simulation  studies using a toy lattice model 
in which
each chain has $N=8$ beads of three types, namely, hydrophobic (H),
polar (P) and charged (see {\em Methods}). 
Our simulations show that the overall assembly of ordered protofilaments and
fibrils occur in three distinct stages.
The smallest time scale is associated with a 
fast "burst phase" during which highly mobile oligomers form.
During this stage there is a distribution of oligomers of varying sizes.
Because we are forced to simulate finite number of chains we cannot quantify
the 
nature of the size distribution.
The second stage is the transformation of the burst phase ....
 into a disordered but
compact oligomer in which
 about half of the interpeptide contacts form. 
It is likely that the conformational transition from 
{\bf S} $ \rightarrow$ {\bf N$^*$} takes
place during this stage as envisioned in the NCC model.
The longest time scale corresponds to the final stage of fibril
formation.
In this stage the large clusters grow by incorporating the small clusters. The
structural transitions here are best described by a dock-lock mechanism
that requires the presence of a template.
Thus, even in the toy model there are complex structural
transitions that take place in each stage of assembly. It appears
that elements of NG, TA, and NCC are operative depending on the stage of
fibril formation.


\section{Methods}


{\em Model.} 
Each chain consists of $N$ connected beads that are confined to the
vertices of a cube.
The simulations are done using
$M$ identical chains with $N$=8. The sequence
 of a chain is +HHPPHH-, where + and - are charged beads.
The assignment of chemical character and the nature of interactions between
the beads should be viewed as a caricature of polypeptide chains, and 
are not realistic
representation of amino acids.
Despite such drastic simplification it has been shown that lattice models are
useful in providing insights into protein folding mechanisms
\cite{Thirumalai02ACP,Shakhnovich91PRL,Socci94JCP}.

The inter- and intra-chain potentials include excluded volume and contact (nearest neighbor)
 interactions. Excluded volume is imposed by the condition that
a lattice site can be occupied by only one bead.
The energy of $M$ chains is
\begin{equation}
E  =  \sum_{l=1}^M \, \sum_{i<j}^N \; e_{sl(i)sl(j)} \delta (r_{ij}-a) +
\sum_{m<l}^M \; \sum_{i,j}^N \; e_{sl(i)sm(j)} \delta (r_{ij}-a) ,
\label{energy_eq}
\end{equation}
where $r_{ij}$ is the distance between residues $i$ and $j$, $a$ is a lattice
 spacing, $sm(i)$ indicates the type of residue $i$ from $m$-th
peptide, and $\delta (0)=1$
 and zero, otherwise.  The first and second terms in Eq. \ref{energy_eq}
 represent intrapeptide and interpeptide interactions, respectively.

The contact energies
between H beads $e_{HH}$ is -1 (in the units of $k_BT$).
 The propensity of polar (including charged) residues to be "solvated"
 is mimicked
using $e_{P\alpha} =$-0.2, 
where $\alpha$= P,+,or -. "Salt-bridge" formation between oppositely charged
beads is accounted for  by a favorable contact energy $e_{+-}=-1.4$. All other contact
 interactions are repulsive. The generic value for
repulsion 
$e_{\alpha\beta}$ is 0.2. For a pair of like-charged beads
the repulsion is stronger, i.e. $e_{++} = e_{--} = (0.7)$. The chains were confined to the vertices
 of the three-dimensional hypercube.
For example, when
$M=10$ the length of is 10$a$. 
Therefore,
 the volume fraction occupied by the peptides is 0.08, and corresponds to the
 concentration of 250 mM. This is about three orders of magnitude denser
than that used in typical experiments.

{\em Simulation details}. Simulations were performed by enclosing $M$
chains in a box with
 periodic boundary conditions. We use Monte Carlo (MC) algorithm to study the
kinetics of amyloid formation.
At the beginning of each MC cycle a peptide is selected at random.
Then one of the two types of MC moves, global or local, is randomly chosen. 
The acceptance probabilities of global and local moves are 0.1 and 0.9,
 respectively.
 Global moves correspond to either translation of a peptide by $a$
 in a randomly chosen direction or rotation by $90^o$ around one
of the randomly
 chosen coordinate axes. The direction of rotation as well as the type of
 global move are selected at random. A local move
\cite{Hilhorst_JCP75} corresponds to
 tail rotation, corner flip,
 and crankshaft rotation. Given the condition that a local move is
accepted of 0.9 probability we used the same relative probabilities
 for selecting the particular types of local moves as described
elsewhere \cite{MSLi_JPCB02}.
We measure time in units of Monte Carlo steps (MCS). The combination of local
and global moves constitutes one MCS.

{\em Structural probes}.
Contacts in the aggregated state (oligomer or fibrils)
are divided into two categories, intrapeptide and interpeptide.
If two non-bonded beads (those that are not covalently linked) of a given 
chain are near-neighbors, then they form an intra-chain contact.
An interpeptide contact in an ordered conformation is one which is 
(i) formed between beads
 belonging to different peptides,
 and (ii) the associated peptide bonds are in the ordered state.
 All interpeptide contacts in the fibril structure satisfy the condition (ii),
 although this is not generally the case for an arbitrary oligomeric structure.
The numbers of intrapeptide and interpeptide fibril contacts
in an arbitrary conformation  are denoted as
$Q_m$ and $Q_f$ with $Q_{m,0}$ and $Q_{f,0}$ being their values
 in the fibril state. In what follows, quantities
 with the subscript “0” correspond to the fibril structure.

In order to probe the growth of the fibril we obtained the distribution
 of fibril clusters in a given oligomer conformation. A fibril cluster is
computed by selecting a pair of fibril contacts and adding adjacent fibril
contacts, whose peptide bonds are parallel or antiparallel to the bonds
associated with original fibril contact pair. The growth of fibril cluster
continues until no more fibril contacts can be added to the cluster in any
 direction. A typical oligomer contains several fibril clusters of different
sizes that are measured by the number of incorporated fibril contacts.
The number of
 fibril contacts in the largest cluster is denoted by $Q_{fc}$. In the fibril
 structure, a single fibril cluster consumes all residues and all chains,
and hence  $Q_{fc} = Q_f$.

We have also computed the number of interpeptide contacts (of any type),
 $C_{out}$, which describes the formation of the aggregated state. Aggregation
 of chains is also monitored by computing the distribution of oligomers.
 An oligomer is defined as a group of aggregated chains. Two oligomers are
 distinct, if none of the chains from one oligomer interacts with any chain
 from the other.  A given multichain conformation may contain several
oligomers and their number, $N_0$, is useful to characterize
the process of aggregation.
In addition, the number of peptides in the largest oligomer $N_p$ is computed.
 As aggregation progresses $N_p$ approaches $M$.

{\em Kinetics of assembly:} To follow the kinetics of aggregation
an initial
 distribution of $M$ random peptide structures is generated,
and equilibrated at high temperature ($T=3.0$) for $10^5$ MCS. The
 resulting distribution of chains is used as a starting point for
initiating
 fibril assembly which begins by quenching the temperature
to $T_s$ ($< 3.0$).
Each MC trajectory starts with a unique distribution of chains. The total
number $N_{MC}$ of MC trajectories for a given $T_s$
varies from 100 to 400.
The first instance, when the fraction of
intrapeptide and interpeptide fibril contacts exceed 0.85
is associated with the first passage time $\tau_{fib,i}$ for fibril assembly 
for a trajectory $i$.
The condition $\gamma (=0.85)$ which is a fraction
of intra- and inter-chain fibril contacts,
was chosen empirically by analyzing numerous MC
trajectories. The mean time of fibril assembly is computed by fitting the
yield of the fibril structure $P_f(t)$ in the pool of $N_{MC}$
independent MC trajectories.

Rapid nucleation of fibril structure was analyzed as follows. For each 
trajectory we considered an interval of $10^6$ MCS immediately preceding
$\tau_{fib,i}$ and computed various quantities associated with
fibril formation as described above.
In addition, within the time interval $\tau_{fib,i}-10^6 < t < \tau_{fib,i}$
we considered the subset of fibril contacts in the largest fibril cluster
$Q_{fc}$, which satisfy two conditions
\cite{Guo97FoldDesign,Klimov98JMB}, namely, (i) that these fibril contacts
 are formed at the time of fibril assembly $\tau_{fib,i}$  and (ii) that
apart from short lived disruptions they remain stable within the
interval ($t,\tau_{fib,i}$). The disruptions of fibril contacts must not exceed
$t=2000$ MCS. The results do not depend on the specific value of
$t$ when it is varied by $\pm$1000 MCS. The fibril contacts satisfying these
two conditions are referred to as "nucleation" fibril contacts and
 their number
is denoted as $Q_{nfc}$.

\section{Results and Discussion}

{\bf Monomeric and fibril structures}

\begin{figure}[ht]
\includegraphics[width=4.00in]{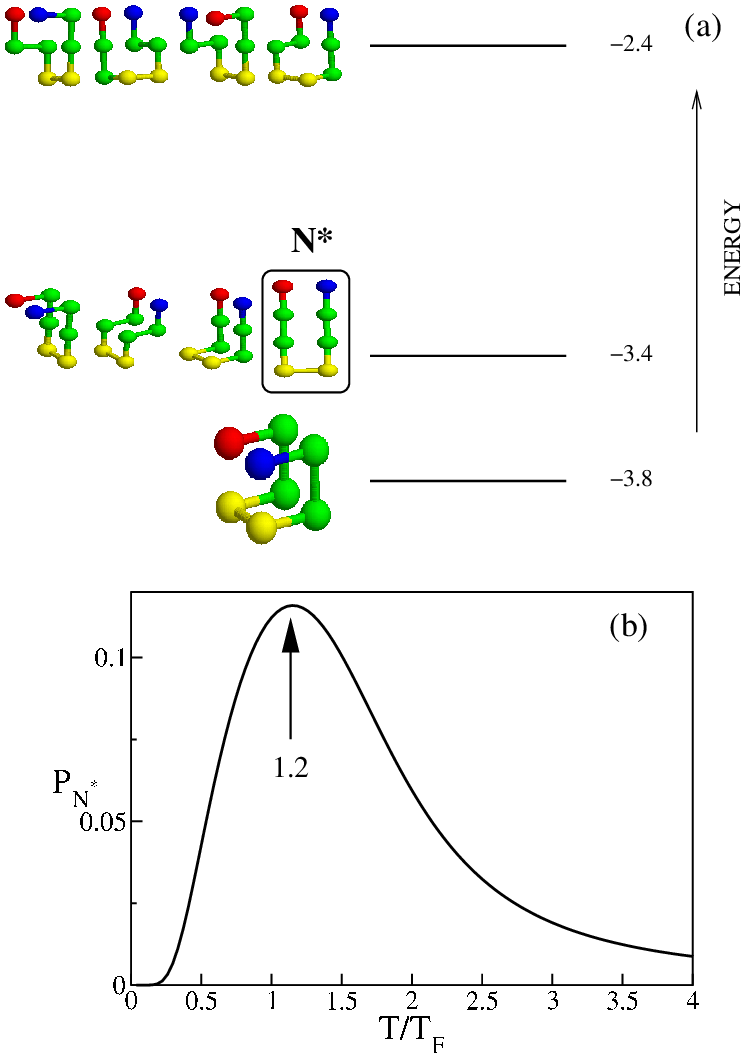}
\caption{}
\label{spectrum_fig}
\end{figure}

{\em Monomer}. 
Exact enumeration of all possible conformations of the monomer
of 8 beads shows
that there are 18 energy levels. Three lowest levels in the
spectrum
are presented in Fig. \ref{spectrum_fig}.
The monomeric native state is compact, and it has the lowest energy $E=-3.8$. 
It should be noted that the conformation of the chain in the fibril state
is not compact
and it belongs to the first excited state 
(label {\bf N$^*$} in Fig.  \ref{spectrum_fig})
which is four-fold degenerate.
Fluctuations in the monomer conformations has
to populate the structure with $E=-3.4$ for oligomerization to start. Such
fluctuations, under condition when the native structure is stable,
can occur spontaneously or through inter-chain interactions. Clearly, 
suppression of fluctuations at low temperatures would slow down the process of
ologimerization.
The toy model captures the well-accepted proposition that aggregation requires
partial unfolding of the native conformation
\cite{Fink98FoldDesign}.

\begin{figure}[ht]
\includegraphics[width=4.in]{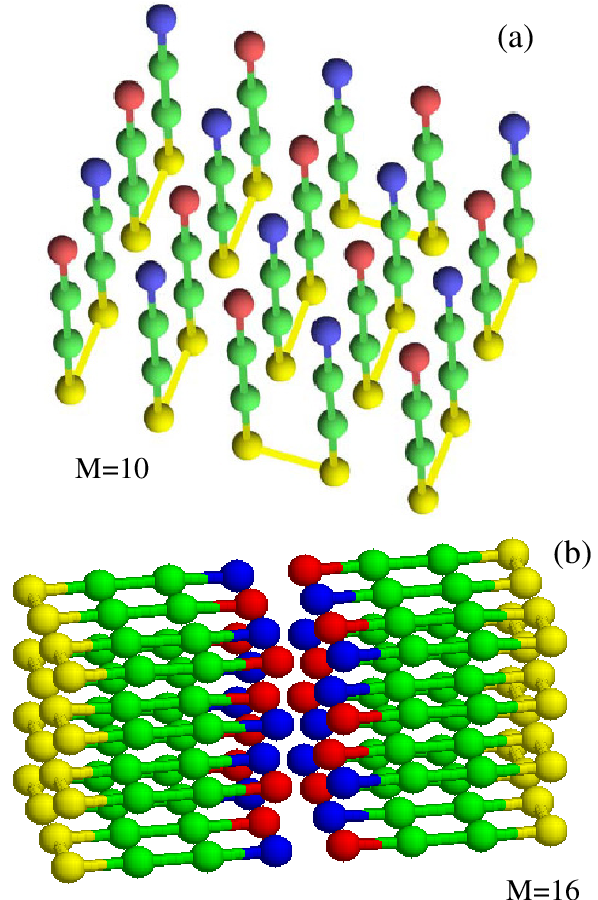}
\caption{}
\label{FibrilStruc_fig}
\end{figure}

{\em Ensemble of peptides}.
When multiple chains are present in the unit cell, aggregation is readily
observed, and in due course they form ordered structures
(Fig. \ref{FibrilStruc_fig}).
Exact enumeration of all conformations for multi-chain systems
 is not possible so that 
the structure of the lowest energy has to be determined using
simulations.
We used the MC
annealing protocol, which allows for the exhaustive conformational search,
to find the lowest energy conformation. In the ordered protofilament
($M=10$) and fibril ($M=16$) structures
the chains adopt an antiparallel arrangement
(Fig. \ref{FibrilStruc_fig}).

The nature of ordering changes depending on $M$, and hence the concentration.
For $M \leq 10$ the chains are arranged in a single layer while for
$M > 10$ the fibril state has a double-layer arrangement 
(Fig. \ref{FibrilStruc_fig}).
Just as noted, using all-atom molecular dynamics simulations
\cite{Tarus06JACS}, the 
organization of chains in the fibril satisfies the principles of
amyloid self assembly ({\bf PASA}) which states fibril structures
are determined by maximizing the
number of salt bridges and hydrophobic contacts
\cite{Tarus06JACS}. In accord with {\bf PASA},
we found that the organization of the lowest energy structure demonstrates
a remarkable order leading to the maximization of favorable electrostatic
and hydrophobic interactions (Fig. 2).
All H (in green) beads located in the
“core” of the fibril are sandwiched between exposed layers of P (in yellow),
and charged beads (in blue and red). More importantly, all peptides adopt
in-registry antiparallel mutual orientation, which implies that for all bond
vectors connecting nearest neighbor pairs of residues $(i,N-i+1)$ and 
$(i+1,N-i)$ $\vec{r}^m_{i,i+1} = - \vec{r}^l_{N-i,N-i+1}$,
where $m$ and $l$ are the peptide indices. The antiparallel
arrangement is enforced by favorable electrostatic interactions.
 Fig. \ref{FibrilStruc_fig} shows that
the nearest neighbors of all negatively charged terminals (in red) are
positively charged beads (in blue).

 For $M=10$,
in all there are 84 interpeptide fibril
contacts and 30 intrapeptide contacts and the entire protofilament
structure in
Fig. \ref{FibrilStruc_fig}a comprises a single layer.
 This implies that a given 
interpeptide
antiparallel in-registry arrangement of chains is translated across the
entire volume of the fibril in all directions. It is interesting that all
intrapeptide contacts are also found in the native
conformation of the monomer (lowest energy conformation in 
Fig. \ref{spectrum_fig}a) and the {\bf N}$^*$ structure
(Fig. \ref{spectrum_fig}a). Due to
different possible distributions of peptides within the volume of a fibril
the lowest energy fibril structure has non-zero entropy.

The fibril contains both interpeptide and intrapeptide
interactions. 
The structure of the monomer in the ordered fibril coincides with one of the 
structures that is higher in energy than the native monomer conformation
(conformation {\bf N$^*$} in Fig. \ref{spectrum_fig}). Because the fibrils are associated with aggregation
 of unfolded structures (here the first excited state in the spectrum of allowed
monomer conformation), it is logical that other morphologies that nucleate 
from different unfolded conformations can form. By scanning the sequences
for $N=8$ we could not produce fibrils starting from high energy monomer 
conformations which highlights one of the limitations of the
lattice model. This observation  suggests
that as long as peptide sequence contains hydrophobic patches
and oppositely charged residues distributed along the sequence the fibril
structure is likely to include a mixture of inter- and intrapeptide interactions
. Combination of inter- and intrapeptide contacts maximizes the number of
hydrophobic and salt bridges thus satisfying the {\bf PASA}.

There are superficial similarities between structures in
Fig. \ref{FibrilStruc_fig} and the model proposed for A$\beta _{1-40}$ whose
sequence is interspersed with charged and hydrophobic residues.
The amyloidgenic A$\beta$ peptide
contains two hydrophobic regions (central hydrophobic cluster and
the C-terminal) as well as charged residues. Proposed fibril model for 
A$\beta_{1-40}$ is based on the assumption that an A$\beta_{1-40}$ monomer
contains a turn, which brings two hydrophobic regions in proximity and
facilitates formation of a salt bridge \cite{Tycko04COSB}.

{\bf Time scales for monomer folding and fibril assembly}

\begin{figure}[ht]
\includegraphics[width=6.00in]{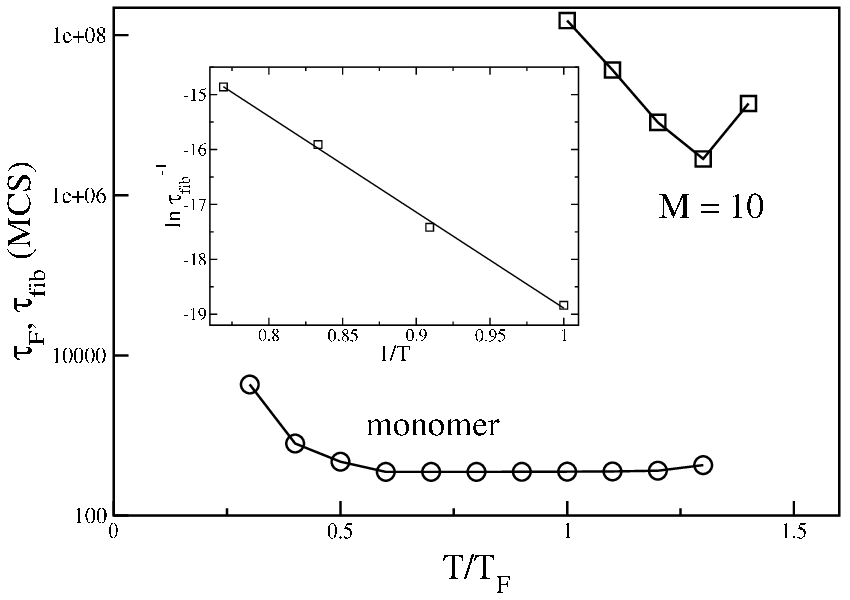}
\caption{}
\label{Fig2_fig}
\end{figure}

The short chain ($N=8$) allows us to compute the times $\tau _F$ for monomer
 folding  as a function of temperature.
The decay of the population of unfolded conformations is best described 
using a single exponential (data not shown) which is characteristic of well
designed sequence. The folding time $\tau _F$ 
is well below $10^3$ MCS (Fig. \ref{Fig2_fig}) over a wide temperature range.
In contrast, the temperature-dependent time for fibril formation,
$\tau _{fib}$, is dramatically different (Fig. \ref{Fig2_fig}). There are
two striking observations about $\tau _{fib}$. 
First, $\tau_{fib}$
is about 4-6 orders of magnitude larger than $\tau_{F}$. Clearly, the sizes
of the monomer and the fibril can cause the vastly greater value of 
 $\tau _{fib}$ compared to  $\tau _F$.
The effect of system size can be roughly rationalized using the approximate
dependence of  $\tau _F$ on $N$ \cite{Thirumalai_JdePhysique95}.
It has been shown that $\tau_F \approx \tau_{F0}e^{1.1\sqrt{N}}$
\cite{MSLi_Polymer04}. Assuming that $\tau_{F0}$
does not change significantly and taking into account that the fibril
in our model is 10 times larger then the monomer size
consideration alone would yield
$\tau _{fib}/\tau _F \sim 10^3$. In addition, formation of fibril
(or protofibrils) also requires collective fluctuation
(formation of nucleus for example)
which requires that several monomers access the {\bf N$^*$}
structure in the first
excited state of the isolated monomer (Fig. \ref{spectrum_fig}). There are barriers associated with such processes that also 
increase $\tau _{fib}$. The relative values 
$\tau _{fib} \sim (10^4 - 10^6) \tau _F$ is not inconsistent
with experimental observations. Typical values of $\tau _F$ for small proteins
is about (1 - 100) ms. Thus, our simulations would suggest 
$\tau _{fib} \sim (10^2 - 10^4)$ sec assuming $\tau _F \sim 10$ ms.

The most striking aspect of Fig. \ref{Fig2_fig} is the dramatic differences in
the $T$-dependence of $\tau _F(T)$ and $\tau _{fib}(T)$. The temperature
independence of $\tau _F(T)$ in the $0.3T_F \leq T \leq 1.3T_F$
is typical of well-designed monomer sequences for which
$T_F \approx T_{\theta}$, where $T_{\theta}$ is
the collapse transition temperature \cite{Klimov_JCP98}.
In contrast, $\tau _{fib}(T)$ changes drastically as $T$ varies. In the narrow
temperature range ($T_F \leq T \leq 1.4T_F$)
$\tau_{fib}$ varies by almost two orders of magnitude.
At the temperature $T \approx 1.3T_F$ (Fig. \ref{Fig2_fig}), when
$\tau _{fib}$ is the smallest,
the native structure is less stable than the unfolded ensemble.
 The structures of
the partially unfolded  conformations at $T \approx 1.3T_F$ shows that
the probability of the "salt bridges" (intramolecular contact between
+ and - beads) being in contact exceeds 0.5. 
At $T \approx 1.3T_F$ there is substantial probability of populating the
aggregation-prone monomer
 (Fig. 1b) that acts as a seed for nucleation and growth.
At $T \approx 1.3T_F$ the fibrils form in the smallest time with 100\% yield
whereas at  $T=T_F$ the yield of the fibril
drops to 0.42 during the simulations lasting of $10^8$ MCS.

The observation that partial
 unfolding of the native state is a necessary condition
 for ologimerization and  fibril growth is consistent with
experimental observations
that many non-homologous protein sequences
 assemble into amyloid fibrils under denaturing conditions 
\cite{Chiti_Nature03}. 
 Although the formation of fibrils is apparently a generic feature of
 polypeptide sequence, our simulations suggest that
for a given sequence there
may be only a narrow window of external conditions that
 favor rapid fibril assembly.
 Besides requiring that the native monomer partially unfolds
for aggregation to begin,
 the denaturing
 conditions must also be relatively mild. Under these conditions
aggregation-prone structures with intramolecular native interactions that 
moderately stable can be populated.
In our model the conformation that nucleate and grow
(Fig. \ref{spectrum_fig}), is homogeneous
which results in a unique fibril structure. Denaturing conditions that 
favor its formation, with intact "salt bridges" results in the most rapid
assembly (Fig. \ref{spectrum_fig}).
In polypeptide chains there may be a collection of conformations that can lead
to fibrils. The differences in fibril morphology is probably linked to the
variations in the initial conformations of the monomer. 

{\bf Fibril assembly occurs in three major stages}

\begin{figure}[ht]
\includegraphics[width=6.00in]{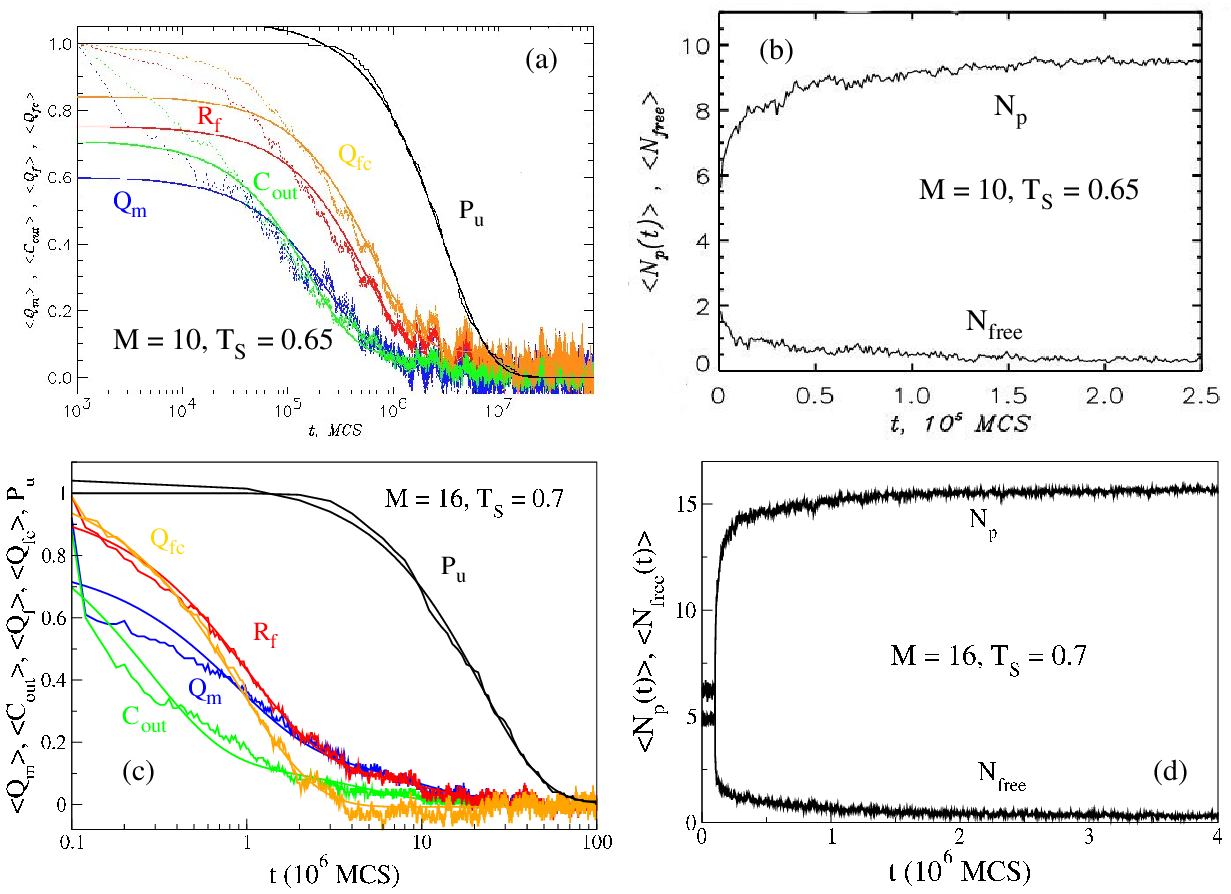}
\caption{}
\label{Fig3_fig}
\end{figure}

{\em Formation of protofilaments:} 
To provide microscopic details of fibril assembly we generated multiple MC 
trajectories for $M=10$ at $T_s = 0.65 = 1.3T_F$ at
 which $\tau _{fib}$ is the smallest
(Fig. \ref{Fig2_fig}).
In all, 100 MC trajectories starting
from random initial conditions were generated. The length of MC trajectories
(8$\times 10^7$ MCS) at $T_s$ was sufficiently long to observe ordered 
structure formation
in each trajectory.
Fig.
\ref{Fig3_fig} 
displays several quantities averaged over 100 trajectories and normalized to
vary from 1 (at $t=0$) to 0 (the equilibrium value).
The averaging over the ensemble of trajectories is indicated by angular
brackets $<…>$. The timescales from exponential fits to these functions
describe the kinetics of fibril formation.
Analysis of the various time dependent quantities and inspection of the
structures sampled enroute to the final fibril gives an intuitive picture of
assembly and growth.

Immediately after temperature quench to $T_s$, the chains
are randomly distributed in the
 unit cell. The numbers of intra- and interpeptide fibril contacts are
 negligible, and there are relatively few interchain interactions.
 The largest oligomer contains, on an average, four chains
 ($N_p$=4). 
Within a short time the inter-chain interactions trigger the formation of
oligomers which represent the growth stage in the route to fibrils.
Fig. \ref{Fig3_fig}b shows that
 the average number of free chains $<N_{free}>$ (those which
 do not make interpeptide contacts) is less than one in 
$\approx 0.03\times 10^6$ MCS or 0.01$\tau_{fib}$. Almost concurrently, the number of
 peptides in the largest oligomer $<N_p(t)>$ exceeds nine.
Thus, already in the initial stage the chains interact and cooperatively form 
fluid-like oligomers.
 Indeed, 
$<N_p(t)>$  grows on the time
 scale of 0.06$\times 10^6$ MCS or 0.02$\tau_{fib}$, and approaches its
 equilibrium value of 9.8. Therefore, virtually all chains are
 incorporated in a single “burst phase” leading to mobile oligomer formation.

 The second
 stage in fibril assembly is associated with the formation of intra- and
 interpeptide interactions, which transforms the mobil oligomer
formed in the first stage, into
 compact disordered oligomer.
During this stage structural rearrangement and conversion from {\bf S} $
 \rightarrow$ {\bf N$^*$}
take place as shown by a number of quantities.
 The intrapeptide fibril contacts $<Q_m(t)>$
 (data in blue in \ref{Fig3_fig}a) are formed on the timescale of 
≈0.1$\tau_{fib}$. On a similar time scale, the number of interpeptide
 contacts $<C_{out}(t)>$ (data in green) approaches the equilibrium value of
 approximately 67. Interestingly, the number of distinct
clusters $<N_{fc}(t)>$ reaches maximum during this stage of fibril
 assembly (data not shown).
 We surmise that the disordered oligomer contains as many as four
 distinct fibril clusters, the largest of which already comprises roughly 
$50\%$ of the entire protofilament.  Fig. \ref{Fig3_fig}a further
 demonstrates that at $t \approx 0.1 \tau_{fib}$ the distribution of the
 volume of fibril clusters extends from predominantly small clusters
 ($Q_{fc} \leq 14$) to larger ones ($15 \leq Q_{fc} \leq 28$). The total
 number of fibril contacts is still relatively small in the disordered
 oligomer ($<Q_f(0.2\tau_{fib})> \approx 30=0.36Q_{fc,0}$). Therefore,
 disordered oligomers are characterized by a nascent single
layer protofilament-like structure (Fig. \ref{FibrilStruc_fig}a) ,
 which emerges in the oligomer volume as a distribution of disjoint fibril
clusters of varying sizes.

The transformation of disordered oligomers to an
ordered structure occurs during the third stage of
 fibril assembly. It follows from Fig. \ref{Fig3_fig}a
 that the timescale for the
 formation of interpeptide fibril contacts $<Q_f(t)>$ is
 0.5$\times 10^6$ MCS or $\approx 0.2\tau_{fib}$  (data in red). Importantly,
on the same time scale the dominant fibril cluster
grows as shown by
$ <Q_{fc}(t)>$ (data in orange). This result indicates that the formation of
 fibril structure occurs via the growth of the largest fibril cluster
at the expense of small clusters.
 The “winner-take-all” scenario of fibril growth is further
described below.
 The number of fibril clusters
 $<N_{fc}(t)>$ decreases to less than
 3 in the time interval of 0.2$\tau_{fib} < t< \tau_{fib}$. On the other hand,
 the maximum in the kinetic distribution of the fibril structure 
among the clusters shifts to
 the right signaling the emergence of large clusters 
($43 \leq Q_{fc} \leq 70$). By assigning weight in proportion
to the size of fibril
clusters
 we find that the dominant fibril cluster comprises
almost the entire fibril
structure. In accord with this conclusion we found that
the fraction of fibril contacts (i.e., the fraction of fibril structure)
in the largest clusters is $43 \leq Q_{fc} \leq 70$ (results not shown).
It is clear that at
$ t>0.4 \tau_{fib}$ more than 80\% of ordered structure is localized in a
 single large fibril cluster.  Because on these time scales
$ <N_{fc}(t>) \approx 2$, the remaining 10 to 20\% of fibril contacts are
 found in a much smaller “satellite” fibril cluster.

 The formation of a dominant
cluster containing the protofilament  also follows from the calculations of thermodynamic quantities.
The thermal averages of the number of fibril contacts $<Q_f>$ and the number
 of fibril contacts in the largest fibril cluster $<Q_{fc}>$ are 52 and 47,
 respectively. Thus, $<Q_{fc}>=0.90<Q_f>$.
After the dominant fibril cluster appears at $t \approx 0.4\tau_{fib}$,
 its further growth and consolidation continues until it reaches its
equilibrium size (about 60\% of all fibril contacts). This kinetic phase can be
 described by additional time scale with small amplitude. Due to this additional
 fibril ordering the final fibril assembly takes place only at
$\tau_{fib} = 3.3\times 10^6$ MCS. Thus, long after the formation of the
largest cluster structural reorganization continues until the ordered
stable fibril forms.
The slow templated-assembly within the large cluster is remimiscent
 of the lock phase.

{\em Mechanism of fibril assembly:}
In order to probe the mechanism of fibril formation
(two-layer structure in Fig. 2b), at $T_s=0.7$,
we generated 100 trajectories with each are being
$10^8$ MCS. The mean time for fibril formation
is  
$\tau _{fib} \approx 2\times 10^7$ MCS. These long runs ensure that 
the fully ordered state is reached in each trajectory. Qualitatively,
 the fibril formation kinetics is the
same as in the $M=10$ case, i.e., it follows three-stage kinetics. However,
there are a few quantitative differences.
In the protofilament formation case the interpeptide
contacts $<C_{out}(t)>$, and intrapeptide fibril contacts $<Q_m(t)>$
(Fig. \ref{Fig3_fig}b) are formed on the same time scale. For $M=16$
(Fig. \ref{Fig3_fig}c) $<C_{out}(t)>$ approaches the value of 0.5 earlier.
Fit of  $<C_{out}(t)>$ using a sum of three exponential functions
gives 
$\tau_1 = 0.15 \times 10^6$ MCS $\approx 0.01 \tau_{fib}$, 
$\tau_2 \approx 10^6$ MCS $\approx 0.05 \tau_{fib}$, and $\tau_3 \approx 11.2 \times 10^6$ MCS $\approx 0.5 \tau_{fib}$.
Thus, $\tau_1$ is a characteristic time scale of
the "burst phase" in which fluid-like clusters form. On this
time scale only a few interpeptide fibril contacts $Q_f$ ($\approx 0.6\%$
of total contacts) are formed and the largest oligomer contains,
 on an average, only five peptides
($N_p$=5).
Using the three-exponential fit and data presented in
Fig. 5a one can show
that the formation of the largest cluster 
occurs on time scale of  
$\approx 0.02 \tau_{fib}$.
The number of peptides
in this cluster approaches 15 (Fig. \ref{Fig3_fig}d) whereas the number of free peptide becomes zero.
Almost simultaneously the number of distinct
fibril clusters $<N_{fc}(t)>$ reaches a maximum 
(data not shown).

The second stage
of fibril assembly, in which the “burst phase” oligomer is transformed
into a compact disordered oligomer, takes place on the times scale
$\tau_2 \approx 0.05 \tau_{fib}$. Due to the larger value
of $M$ this time is larger than
for $M=10$ .
At this stage 50\% of equilibrium values of the intra- ($Q_m$) 
and interpeptide ($Q_f$) fibril contacts are formed. Contrary to
the $M=10$ case,
fibril contacts in the largest cluster $Q_{fc}$ are formed earlier
than total $Q_f$. This is probably due to increasing role
of the satellite clusters as the number of monomers increases.
On long time scales we have more than two 
and less than two
such clusters for $M=16$ and 10, respectively. 
The "winner-take-all scenario" is also valid for the
$M=16$ system because for $t > 0.2\tau_{fib}$ the largest cluster contains
$\approx 75\%$ of fibril contacts.
These observations are made quantitative using the dependence of
$\tau _{fib} \sim M$ (see below).

As seen from Fig. \ref{cont_fib_total_t07_fig}a, the three exponentials
($f(t) = f_0 - f_1\exp(-t/\tau_1) - f_2\exp(-t/\tau_2) - f_3\exp(-t/\tau_3)$)
fit the data well (dashed line). Here, we have three different time scales
$\tau_1 \approx 0.17 \times 10^6, \tau_2 \approx 1.24 \times 10^6$
and $\tau_1 \approx 12.18 \times 10^6$ MCS (the partition of these phases
is $f_1 \approx 0.19, f_2 \approx 0.46$ and
$f_3 \approx 0.1$). 
Experiments \cite{Esler00Biochem} on the fibril growth kinetics of
A$\beta$-peptides, that is fit using a sum of two exponential functions,
have been interpreted in term of templated-assisted "dock-lock" mechanism.
From the perspective of the present studies
we conclude that such a mechanism is probably valid during the second and
third stages of fibril growth. The lock phase during which in-registry
arrangement of the chains takes place, clearly occurs only during the last part
of stage three in the fibril growth process.
 The early stages of growth reveal a much more complex set 
of events  in which physical process described in NG and NCC are manifested 
(see also the {\em Concluding remarks}).

{\bf Dependence of fibril formation time on number of monomers.}

In order to obtain the dependence of $\tau_{fib}$ on number of monomers, we fixed
the monomer concentration and computed $\tau_{fib}$ for each system at $T_s$.
The fibril formation time scales linearly with the number of monomer
(Fig. 5b), $\tau_{fib} \sim M$ but with
different slopes for $M \leq 10$ and $M > 10$. This is probably related to
difference between protofilament and fibril formation
(see Fig. \ref{FibrilStruc_fig}{\em a} and \ref{FibrilStruc_fig}{\em b})
The linear dependence of $\tau_{fib}$ on $M$ supports the template-assisted
mechanism in which monomers are added one by one to preformed ordered structures
(protofibrils or fibrils)
provided the number of these monomers exceeds the size of critical nucleus.
Thus, the linear dependence characterizes growth only during the
late stages of ordered assembly.
Our results agrees with experimental findings of Kowalewski and Holtzman
\cite{Kowalewski99PNAS} who studied aggregation of Alzheimer's $\beta$-amyloid
peptides on hydrophilic mica and hydrophobic graphite surfaces as well as 
with the results obtained by Collins {\em et al.} \cite{Collins04PLOSBIOL}
for the amyloidogenic yeast prion protein Sup35. 

Interestingly, the dependence of $\tau_{fib}$ on $M$ for such a complicated
process as fibril assembly seems to follow the well-known
Lifshitz-Slyzov law. 
Since $ M \sim L^3$, where $L$ is a typical size
of the ologimer, we obtain 
$\tau_{fib} \sim L^{1/3}$ which is the Lifshitz-Slyzov law
 \cite{Lifshitz61JPCS} describing the growth of a cluster in a supersaturated
solution.  The finding in Fig. 5b further supports the "winner-take-over"
scenario for oligomer growth because the Lifshitz-Slyzov law is based on the
assumption that the largest cluster grows at the expense of smaller ones.


\section{Concluding remarks}

We have used a lattice model to elucidate the generic features of
fibril assembly mechanisms in proteins.
Using this toy model many aspects of
the transitions from the monomer to fully formed fibrils
can be monitored. Examination of the kinetics of the assembly process
reveals that several aspects of   complex set of transitions seen
in the simple model is also qualitatively observed in experiments.


\begin{enumerate}
\item

The ordered fibrils form as the number of chains become greater than
critical value. In our system we
find that for $M = 16$ a stable two layer fibril is formed which is
perhaps the minimum replicating unit in
the infinite fibril. For smaller $M$ (Fig. 2a) ordered protofilaments
are the lowest energy conformation. It
is likely that there are substantial internal rearrangements of the
chains as the number of monomers increases
so that a stable fibrils can be populated. Although, we did not carry
out systematic calculations to infer the
size of the critical nucleus it appears both from the temperature dependence
of protofilament formation as well as the ease of fibril production
for $M = 16$ that the size of the nucleus has to be
less than 10.

\item
The kinetics of fibril formation occurs broadly in three distinct
stages. In the initial stage, the chains rapidly partition into
clusters of varying sizes. Because of finite size limitations we are
unable to determine the precise distribution of cluster sizes.  The chains
within each cluster is mobile and fluid-like. There are, in all
likelihood, substantial conformational fluctuations within each
cluster.
In the second stage the chains in the clusters form a number of intra-
and inter-chain contacts that leads to the disordered oligomers.
During this stage bigger clusters grow at the expense of smaller ones.
In the process protofibrils in which many peptide adopt the
eventual conformation
in the fibrils form. In the third post-nucleation stage the chains add
to the largest (single) cluster. In this stage, which is captured in
experiments, the addition of
a monomer occurs by a lock-dock mechanism.  Thus, a cascade of events
starting from conformational fluctuations in the monomer that populate
the
aggregation-prone conformation (Fig. 1a) through a series of
inter-peptide interaction-driven conformational changes results in
fibril assembly.

\item
The growth kinetics depends on the depth of quench $\Delta T = (T_i -
T_s)$, where $T_i$ is the initial temperature at which the
chains are brought to equilibrium.  When the depth of quench is large then there appears to be
a lag-time before the fibrils are populated. In this case the ordered
structures  form in a highly cooperative manner. In contrast, when the
growth process
is initiated by equilibrating the monomers at the final growth
temperature ($\Delta T = 0$) then the fibril growth occurs in a
continuous manner
and is  less cooperative (Fig. 6 ). 
Because the aggregation-prone structure is unique  in the
toy model we do not observe variations in the morphology of the
final fibril structure.  This is surely an artifact of the lattice model.

\item
The temperature dependence of $\tau_{fib}^{-1}$ for
$M = 10$ shows  Arrhenius behavior with $\tau_{fib}^{-1} \sim
\exp(-E_A/k_{\rm B}T)$ (see inset in Fig. 3). This
is in qualitative agreement with experiments
\cite{Sabate05IJBMM,Kusumoto98PNAS}.
In addition, collective rearrangement of
several chains
from the \textbf{S} to the \textbf{N$^*$} structure that occurs within
the oligomer becomes slower
at low temperatures. These two factors contribute to the barrier that
leads to substantial increase in $\tau_{fib}$ as $T$ is lowered.

\item
The mechanism of assembly of fibrils even in this toy model is highly
complex. While the overall growth kinetics can can be
summarized using a three stage growth the events that transpire in the
distinct stages involve large structural transitions.  In the initial
"burst phase" loosely bound clusters form in which the chains are
essentially "non-interacting". In the second stage stable clusters
with
considerable inter-particle interactions form. There is a distribution
of oligomers. Due to finite size of the simulations the nature of
distribution is unclear.
It is within these oligomers, in which the chains are in a mixture of
\textbf{S}-like and the aggregation prone \textbf{N$^*$}-like states,
the
conversion from \textbf{S} to \textbf{N$^*$} takes place. These
transitions result in formation of large-enough
ordered oligomers 
that can serve as
templates for conversion of  additional monomers or oligomers to form
mature fibrils. It is the last stage that is best described by the
dock-lock mechanism.

\item
Strikingly, the growth of mature fibrils in the third stage occurs by
the Lifschitz-Slyazov mechanism in which the largest clusters grow at
the
expense of smaller ones.  The proposed mechanism supports the physical
picture that \textbf{S} $\rightarrow$ \textbf{N}$^*$ transition occurs
either in the oligomers (NCC model) or upon addition to preformed
ordered template (dock-lock mechanism).     Thus, we find that
elements
of the three models (NG, TA, and NCC) are found in each assembly
stage. This conclusion also supports a detailed study of fibril growth
in off-lattice model of poly-alanine \cite{Nguyen05JBioChem} in which
multiple routes to fibril formation was found even in the final stages
of incorporation
of ordered structures or disordered monomers.   Finally, the proposed
Lifschitz-Slyazov growth law strongly suggests that seeding with
preformed fibrils
should lead to rapid growth because such large structures can
incorporate disordered oligomers on time scales that vary linearly with peptide
concentration.
\end{enumerate}

Acknowledgment: MSL and DKK thank the hospitality of colleagues at IPST where
this work was initiated. This work was supported by the Polish KBN
grant No 1P03B01827.

\clearpage

\bibliographystyle{pnas}
\bibliography{FibLatt}

\clearpage

\begin{center}
{\bf Figure captions}
\end{center}

\vskip 5 mm

\noindent {\bf FIGURE 1.}
(a) Energies and structures of some of the conformations
of the monomer using exact numeration.
Hydrophobic, polar, positively and negatively charged
beads are shown in green, yellow, blue, and red, respectively.
 There are a total
1831 possible conformations that are  spread among
18 possible energy values. The non-degenerate native conformation
is separated from degenerate higher energy conformations. The
structure enclosed in the box is the one that the chain adopts in the  fibril,
and is referred to as {\bf N$^*$}.
The second highest energy structures are also four-fold degenerate.
(b) The probability $P_{N^*}$ of populating the aggregation-prone
 structure {\bf N}$^*$ as a function of $T$.
The arrow indicates the temperature at which $P_{N^*}$ is maximum.

\vskip 5 mm

\noindent {\bf FIGURE 2.}
a) The lowest energy structure
for ten monomers ($M=10$). The chains are arranged in
an antiparallel manner. The structure of the monomer is the same as the
{\bf N}$^*$ conformation in Fig. 1a. Beads are colored in the same
manner as in Fig. 1a.
This single layer structure is a protofilament.  b) The double
layer structure of $M=16$.
As in fibrils of polypeptides the "$\beta$-sheet like" monomers are arranged
perpendicular to the fibril axis which lies parallel to the
"salt-bridge" plan (contact between blue and red).
Thus, the protofilament and the fibril are stabilized by hydrophobic
interactions and salt-bridges.

\vskip 5 mm

\noindent {\bf FIGURE 3.}
The temperature dependence of monomeric folding time $\tau _F$
(open circles) and the time for protofilament
assembly $\tau _{fib}$ (squares) for $M=10$.
 Temperature is given in the units of the
 monomer folding temperature $T_F=0.5$. This value of $T_F$ is obtained using
the condition
$<Q(T_F)>=0.5$, where $<Q(T_F)>$ is the fraction of native contacts.
The inset shows $\tau_{fib}^{-1}$ at low
temperatures as a function
of $1/T$ for $M = 10$.

\vskip 5 mm

\noindent {\bf FIGURE 4.}
(a)Time dependence of structural quantities probing the formation of
 fibril structure. The number of intrapeptide fibril contacts $<Q_m(t)>$,
 the number of interpeptide contacts $<C_{out}(t)>$, the number of fibril
 contacts $<Q_f(t)>$, and the number of fibril contacts in the largest fibril
 cluster $<Q_{fc}(t)>$ are shown in blue, green, red, and orange, respectively.
 The data are averaged over 100 trajectories, and smooth lines represent the
 biexponential fits to the data. The fraction of trajectories in which the
fibril structure is still not reached, $P_u(t)$, is shown in black. (b)
The time dependence of the number of peptides in the largest oligomer
$<N_p(t)>$ and the number of free peptides $<N_{free}(t)>$.
(c) Same as in (a) except the results for $M=16$. (d) Same as
(b) but for $M=16$.

\vskip 5 mm

\noindent {\bf FIGURE 5.}
(a) Time dependence of the fraction of fibril structure for $M=16$
and $T=0.7$. The dashed curve corresponds to fit of the
simulated data using a sum of
three exponentials.
(b) Dependence of $\tau _{fib}$ as a function of $M$. The change in the slope
for $M > 10$ corresponds to the transition from profilament (single
layer) to fibrils (double layer).

\vskip 5 mm

\noindent {\bf FIGURE 6.}
Dependence of the fibril fraction $f(t)$ for the pentamer
($M=5)$ at $T=0.4$ ($< T_F$).
The initial conformations for the high $T$-quench were generated by
equilibrating the pentamer for $10^5$ MCS at $T=2.0$. Subsequently assembly
of the fibril fragment is initiated by quenching the temperature to $T_s = 0.4$.In the low-$T$ quench regime the initial
configurations were generated in the same way but equilibration
was done
at $T=0.4$.
Typical snapshots at various times during the fibril growth are shown.

\clearpage

\clearpage

\clearpage

\clearpage

\clearpage

\begin{figure}[ht]
\includegraphics[width=5.00in]{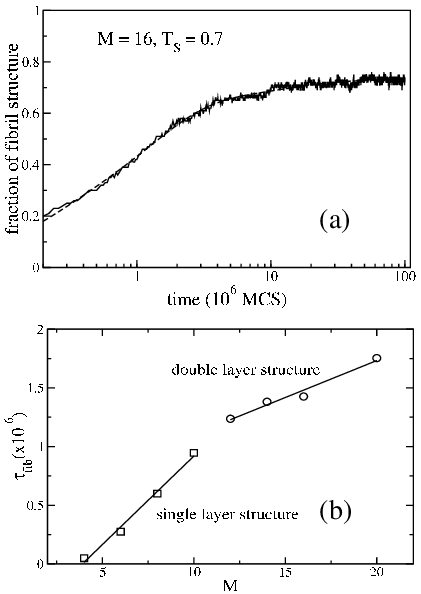}
\caption{}
\label{cont_fib_total_t07_fig}
\end{figure}

\clearpage

\begin{figure}[ht]
\includegraphics[width=6.00in]{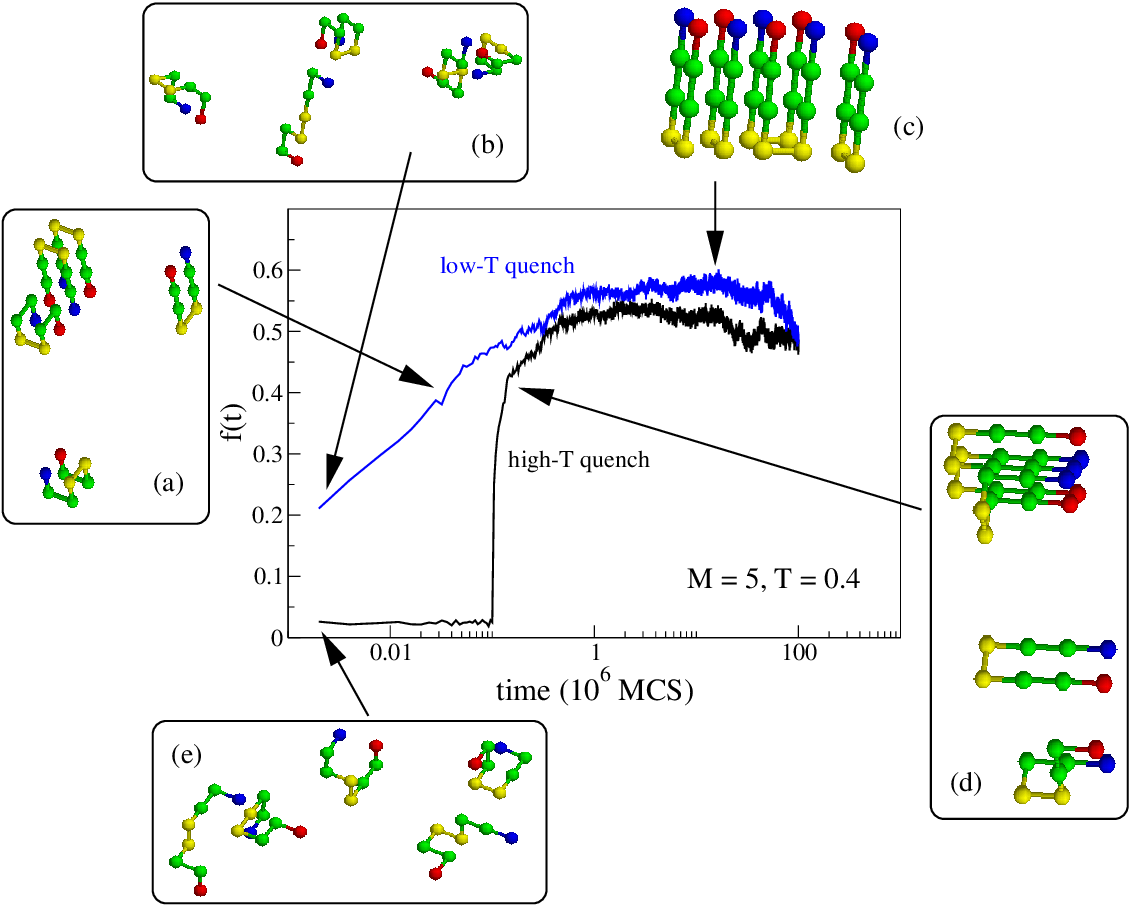}
\caption{}
\label{cont_fib_5mer_t04_fig}
\end{figure}

\end{document}